\documentstyle[preprint,revtex,eqsecnum]{aps}
\tightenlines
\begin{document}
\draft

\preprint{SLAC--PUB--5898}
\medskip
\preprint{September, 1992}
\medskip
\preprint{T/E}

\begin{title}
Second Order Power Corrections in the Heavy Quark\\
Effective Theory\\
II. Baryon Form Factors
\end{title}

\author{Adam F. Falk and Matthias Neubert}
\begin{instit}
Stanford Linear Accelerator Center\\
Stanford University, Stanford, California 94309
\end{instit}

\begin{abstract}

The analysis of $1/m_Q^2$ corrections of the previous paper is extended
to the semileptonic decays of heavy baryons. We focus on the simplest
case, the ground state $\Lambda_Q$ baryons, in which the light degrees
of freedom are in a state of zero total angular momentum. The
formalism, while identical in spirit, is considerably less cumbersome
than for heavy mesons. The general results are applied to the
semileptonic decay $\Lambda_b\to\Lambda_c\,\ell\,\nu$. An estimate of
the leading power corrections to the decay rate at zero recoil, which
are of order $1/m_Q^2$, is presented. It is pointed out that a
measurement of certain asymmetry parameters would provide a direct
measurement of $1/m_Q^2$ corrections. Finally, it is shown how the
analysis could be extended to include excited heavy baryons such as the
$\Sigma_Q$ and the $\Sigma_Q^*$.

\end{abstract}

\centerline{(Submitted to Physical Review D)}
\newpage
\narrowtext

\section{Introduction}
\label{sec:1}

In the previous paper \cite{Mesons} (hereafter referred to as Ref.~I),
we have developed the formalism for including in the heavy quark
effective theory (HQET) terms in the mass expansion of order $1/m_Q^2$.
That paper focused on the case of the ground state pseudoscalar and
vector mesons. Here we extend the analysis to the case of the heavy
baryons, in particular the spin-${1\over2}$ $\Lambda_Q$. It turns out
that the formalism is far less cumbersome than for the heavy mesons.
The structure of the previous paper may be taken over almost in its
entirety to the baryons, but with the number of invariant form factors
considerably reduced.  Hence to avoid redundancy we will abbreviate
considerably those aspects of the presentation which are common to the
two cases, and concentrate instead on features which distinguish the
baryons from the mesons.  In Sec.~\ref{sec:2} we discuss the Lagrangian
of HQET and the expansion of the baryon masses. Sec.~\ref{sec:3}
reviews the form of baryon matrix elements in the $m_Q\to\infty$ limit
and the corrections of order $1/m_Q$. In Sec.~\ref{sec:4} we present
the extension of this analysis to order $1/m_Q^2$. Some
phenomenological applications of our results to semileptonic decays of
the $\Lambda_b$ are discussed in Sec.~\ref{sec:5}, while
Sec.~\ref{sec:6} contains a discussion of excited baryons.  In
Sec.~\ref{sec:7} we provide a brief summary.

For the sake of simplicity, we shall completely ignore radiative
corrections in this paper. In particular, we omit the $\mu$-dependence
of the universal form factors of HQET, and ignore the short-distance
coefficients in the expansion of the currents. All these effects would
not change the structure of the heavy quark expansion, but they would
complicate considerably the presentation. As discussed in detail in
Ref.~I, renormalization effects may be incorporated straightforwardly
into our general formalism in a perturbative way.

\section{The Lagrangian of the Effective Theory}
\label{sec:2}

The heavy quark effective theory  provides an expansion of strong
matrix elements in inverse powers of the mass of a heavy quark
\cite{Volo,Isgu,Eich,Lepa,Grin,Geor,Falk,Main}. It is useful when one
considers external states containing a single heavy quark, dressed by
light degrees of freedom to make up a color singlet hadron. HQET is
constructed by redefining the field operator $Q(x)$ of a heavy quark in
such a way that the heavy quark part of the QCD Lagrangian can be
expanded in powers of $1/m_Q$. This expansion is independent of the
nature of the hadronic states one wants to describe. Hence the field
redefinition and the construction of the effective Lagrangian and the
effective heavy quark currents are the same as described in Ref.~I.

In brief, then, there are two objects which one must expand to
construct HQET.  The first is the QCD Lagrangian.  In the limit
$m_Q\rightarrow\infty$, the heavy quark field $Q(x)$ is replaced by the
velocity-dependent field
\begin{equation}
   h(v,x)=e^{im_Qv\cdot x}\,P_+\,Q(x) \,,
\end{equation}
where $P_+=\case1/2(1+\rlap/v)$ is a positive energy projection
operator.  The effective Lagrangian for the strong interactions of a
heavy quark becomes \cite{Geor,Mann,Korn}
\begin{equation}\label{laghqet}
   {\cal L}_{\rm HQET} = \bar h\, iv\cdot D\, h \,,
\end{equation}
where $D^\alpha = \partial^\alpha - ig_sT_aA_a^\alpha$ is the
gauge-covariant derivative.  This is corrected by an infinite series of
terms involving higher dimension operators, which are suppressed by
inverse powers of $m_Q$:
\begin{equation}\label{lpower}
   {\cal L}_{\rm power} = {1\over2m_Q}{\cal L}_1 +
   {1\over4m_Q^2}{\cal L}_2 + \cdots\,.
\end{equation}
The terms in ${\cal L}_{\rm power}$ are treated as ordinary
perturbations of the Lagrangian ${\cal L}_{\rm HQET}$.  Omitting
operators which vanish by the equations of motion, the first and second
order terms are \cite{FGL,Luke,Lee}
\begin{eqnarray}\label{L1}
   {\cal{L}}_1 &=& \bar h\,(i D)^2 h
    + Z\,\bar h\,s_{\alpha\beta} G^{\alpha\beta} h \,, \nonumber\\
   && \\
   {\cal{L}}_2 &=& Z_1\,\bar h\,v_\beta i D_\alpha G^{\alpha\beta} h
    + 2 Z_2\,\bar h\,s_{\alpha\beta} v_\gamma i D^\alpha
    G^{\beta\gamma} h \,, \nonumber
\end{eqnarray}
where $s_{\alpha\beta}=-{i\over 2}\sigma_{\alpha\beta}$, and
$G^{\alpha\beta}=[i D^\alpha,i D^\beta]=i g_s T_a G_a^{\alpha\beta}$ is
the gluon field strength. Expressions for the renormalization factors
have been given in Ref.~I. It is necessary to perform a similar
expansion of the heavy quark currents which mediate the weak decays of
heavy hadrons.  In the full theory these currents are of the form $\bar
Q'\,\Gamma\,Q$.  At tree level in the effective theory the expansion
takes the form
\begin{eqnarray}\label{Jexp1}
   \bar Q'\,\Gamma\,Q &\to& \bar h'\,\Gamma\,h
    + {1\over 2 m_Q}\,\bar h'\,\Gamma\,i\,\rlap/\!D\,h
    + {1\over 2 m_{Q'}}\,\bar h'\,(-i\overleftarrow{\,\rlap/\!D})\,
    \Gamma\,h \nonumber\\
   &&+ {1\over 4 m_Q^2}\,\bar h'\,\Gamma\,\gamma_\alpha v_\beta
    G^{\alpha\beta} h - {1\over 4 m_{Q'}^2}\,
    \bar h'\,\gamma_\alpha v'_\beta G^{\alpha\beta} \Gamma\,h
    \nonumber\\
   &&+ {1\over 4 m_Q m_{Q'}}\,\bar h'\,(-i\overleftarrow{\,\rlap/\!D})
      \,\Gamma\,i\,\rlap/\!D\,h + \cdots\,.
\end{eqnarray}
A more complete form of the expansion, which allows for the inclusion
of radiative corrections, is given in Ref.~I.

The eigenstates of ${\cal L}_{\rm HQET}$ differ from those of the full
theory in the baryon sector in the same way as in the meson sector. The
latter case was discussed in some detail in the previous paper. For the
spin-${1\over2}$ $\Lambda_Q$ baryon the situation is in fact simpler,
because the light degrees of freedom carry no angular momentum and
hence there is no spin symmetry violating mass splitting.  We expand
the mass of the physical $\Lambda_Q$ as $m_\Lambda=m_Q+\bar\Lambda+
\Delta m_\Lambda^2/2m_Q+\cdots$.  The mass of the $\Lambda_Q$ in the
strict $m_Q\to\infty$ limit is given by $M\equiv m_Q+\bar\Lambda$; the
next term in the series represents the leading correction to this
quantity.  Fixing, as usual, the heavy quark mass $m_Q$ so that there
is no residual mass term \cite{AMM} in the Lagrangian (\ref{laghqet}),
the parameter $\bar\Lambda$ is well defined and controls the phase of
the effective heavy baryon state:
\begin{equation}\label{space}
   |\Lambda(x)\rangle_{\rm HQET} = e^{-i\bar\Lambda v\cdot x}
   |\Lambda(0)\rangle_{\rm HQET} \,.
\end{equation}
Note that $\bar\Lambda$ as defined here is {\it not\/} the same as the
analogous parameter $\bar\Lambda$ defined for the heavy mesons. In
order to make clear the parallels with the analysis for mesons given in
Ref.~I, and in order to avoid a further proliferation of nomenclature,
we will sometimes use the same (or similar) names for parameters and
form factors appearing in the description of heavy mesons and baryons.
However, under no circumstances should there be confusion that these
form factors are at all related.

In the rest frame of the $\Lambda_Q$, the mass shift
$\Delta m_\Lambda^2$ is given by
\begin{equation}\label{dm2}
   \Delta m_\Lambda^2 = {\langle \Lambda(v,s) |\,(- {\cal{L}}_1)\,|
   \Lambda(v,s)\rangle \over\langle \Lambda(v,s) |\,h^\dagger h\,|
   \Lambda(v,s)\rangle} \,.
\end{equation}
The matrix elements which appear in the numerator of
(\ref{dm2}) are restricted by Lorentz invariance to take the form
\begin{eqnarray}\label{lamdef}
   \langle \Lambda |\,\bar h\,(i D)^2 h\,| \Lambda\rangle
   &=& 2 m_\Lambda\lambda, \nonumber \\
   \langle \Lambda |\,\bar h\,s_{\alpha\beta} G^{\alpha\beta} h\,
   | \Lambda\rangle &=& 0 \,.
\end{eqnarray}
Vector current conservation implies that the matrix element in the
denominator equals $2 m_\Lambda$. We thus find $\Delta m_\Lambda^2=
-\lambda$.  At this order in the heavy quark expansion, then,
$\bar\Lambda$ and $\lambda$ are the fundamental mass parameters of the
effective theory. They are independent of $m_Q$ and of the
renormalization scale $\mu$. Unfortunately, these parameters cannot be
measured directly. While one may na\"\i vely estimate
$\bar\Lambda\approx700\,{\rm MeV}$ from the constituent quark model,
little is known about the higher order correction $\lambda$.

\section{Baryon Form Factors in the Effective Theory}
\label{sec:3}

Consider the semileptonic decay of a spin-${1\over2}$ baryon $\Lambda$
containing heavy quark $Q$ of mass $m_Q$, to a spin-$1\over2$ baryon
$\Lambda'$ containing heavy quark $Q'$ of mass $m_{Q'}$. This
transition is governed by the hadronic matrix elements of the flavor
changing vector and axial vector currents. They are conventionally
parameterized in terms of six form factors $f_i$ and $g_i$, defined by
\begin{eqnarray}
   \langle \Lambda'(p',s')|\,\bar Q'\gamma^\mu Q\,|\Lambda(p,s)\rangle
   &=& \bar u_{\Lambda'}(p',s') \Big[ f_1\,\gamma^\mu
    -i f_2\,\sigma^{\mu\nu} q_\nu + f_3\,q^\mu \Big] u_\Lambda(p,s) \,,
     \nonumber \\
   && \\
   \langle \Lambda'(p',s')|\,\bar Q'\gamma^\mu\gamma^5 Q\,
   |\Lambda(p,s)\rangle &=& \bar u_{\Lambda'}(p',s') \Big[
    g_1\,\gamma^\mu - i g_2\,\sigma^{\mu\nu} q_\nu
    + g_3\,q^\mu \Big] \gamma^5\,u_\Lambda(p,s) \,,
    \nonumber
\end{eqnarray}
where $q^\mu=p^\mu-p'^\mu$ is the momentum transfer to the leptons. For
heavy baryons it is convenient to replace this with a parameterization
in terms of the velocities of the initial and final baryons. We thus
define an equivalent set of form factors by
\begin{eqnarray}
   \langle \Lambda'(v',s')|\,\bar Q'\gamma^\mu Q\,|\Lambda(v,s)\rangle
    &=& \bar u_{\Lambda'}(v',s') \Big[ F_1\,\gamma^\mu + F_2\,v^\mu
     + F_3\,v'^\mu \Big] u_\Lambda(v,s) \,, \nonumber\\
   && \\
   \langle \Lambda'(v',s')|\,\bar Q'\gamma^\mu\gamma^5 Q\,
   |\Lambda(v,s)\rangle &=& \bar u_{\Lambda'}(v',s') \Big[
    G_1\,\gamma^\mu + G_2\,v^\mu + G_3\,v'^\mu \Big] \gamma_5\,
    u_\Lambda(v,s) \,.\nonumber
\end{eqnarray}
Here $u_\Lambda(p,s)$ and $u_\Lambda(v,s)$ are the same spinors, and
are normalized to the physical mass $m_\Lambda$:
\begin{equation}
   \bar u_\Lambda(v,s)\,u_\Lambda(v,s) = 2 m_\Lambda \,.
\end{equation}
While the form factors $f_i$ and $g_i$ are conventionally written in
terms of the invariant momentum transfer $q^2$, it is more appropriate
to consider $F_i$ and $G_i$ as functions of the kinematic variable
$w=v\cdot v'$, which measures the change in velocity of the heavy
baryons. Using the fact that the spinors are eigenstates of the
velocity, $\rlap/v\,u_\Lambda(v,s)=u_\Lambda(v,s)$, one can readily
derive the relations among these sets of form factors. They are
\begin{eqnarray}\label{fsandgs}
   f_1 &=& F_1 + (m_\Lambda + m_{\Lambda'})
    \bigg( {F_2\over 2 m_\Lambda} + {F_3\over 2 m_{\Lambda'}} \bigg)
    \,, \nonumber \\
   f_2 &=& - {F_2\over 2 m_\Lambda} - {F_3\over 2 m_{\Lambda'}}
    \,, \nonumber \\
   f_3 &=& {F_2\over 2 m_\Lambda} - {F_3\over 2 m_{\Lambda'}}
    \,, \nonumber \\
   && \\
   g_1 &=& G_1 - (m_\Lambda - m_{\Lambda'})
    \bigg( {G_2\over 2 m_\Lambda} + {G_3\over 2 m_{\Lambda'}} \bigg)
     \,, \nonumber \\
   g_2 &=& - {G_2\over 2 m_\Lambda} - {G_3\over 2 m_{\Lambda'}}
    \,, \nonumber \\
   g_3 &=& {G_2\over 2 m_\Lambda} - {G_3\over 2 m_{\Lambda'}} \,.
    \nonumber
\end{eqnarray}

Let us now review the analysis of the baryon form factors in HQET
\cite{IWbar,Gebar,MRR,GGW}. This will allow us to outline the procedure
and to set up our conventions in such a way that the extension to the
next order becomes straightforward. At each order in the heavy quark
expansion, one writes the contributions to $F_i$ and $G_i$ in terms of
universal, $m_Q$-independent form factors, which are defined by matrix
elements in the effective theory. At leading order, one needs the
matrix elements of the first operator on the right-hand side of
(\ref{Jexp1}) between baryon states in the effective theory. They have
the structure \cite{Geor,MRR}
\begin{equation}\label{lowest}
   \langle \Lambda'(v',s')|\,\bar h'\,\Gamma\,h\,|\Lambda(v,s)\rangle
   = \zeta(w)\,\,\overline{\cal{U}'}(v',s')\,\Gamma\,
   \,{\cal{U}}(v,s) \,,
\end{equation}
where $\zeta(w)$ is the Isgur-Wise function for $\Lambda$ baryon
transitions, and ${\cal{U}}(v,s)$ denotes the spinor for a heavy baryon
in the effective theory. It is normalized to the effective mass $M=m_Q
+ \bar\Lambda$ of the state in HQET,
\begin{equation}
   \overline{\cal{U}}(v,s)\,\,{\cal{U}}(v,s) = 2 M \,,
\end{equation}
and is thus related to the spinor of the physical state by
\begin{equation}\label{spinors}
   {\cal{U}}(v,s) = Z_M^{-1/2}\,u(v,s) \,, \quad
   Z_M = {m_\Lambda\over M} = 1 - {\lambda\over 2 m_Q^2} + \cdots \,.
\end{equation}
At order $1/m_Q^2$ in the heavy quark expansion we will have to include
this factor.

 From (\ref{lowest}) one can immediately derive expressions for the
baryon form factors in the infinite quark mass limit. One finds
$F_1=G_1=\zeta(w)$ and $F_2=F_3=G_2=G_3=0$. One can then use the
conservation of the flavor-conserving vector current to derive the
normalization of the Isgur-Wise form factor at zero recoil \cite{Isgu}.
 From
\begin{equation}\label{normalize}
   \langle\Lambda(v,s)|\,\bar Q\,\gamma^0\,Q\,|\Lambda(v,s)\rangle
   = 2 m_\Lambda v^0\,
\end{equation}
it follows that
\begin{equation}\label{CVC}
   \sum_{i=1,2,3} F_i(1) = 1 \,, \qquad (m_\Lambda = m_{\Lambda'})
\end{equation}
which implies the normalization condition $\zeta(1)=1$. From here on we
will omit the velocity and spin labels on the states and spinors. It is
to be understood that unprimed objects refer to $\Lambda$ and depend on
$v$ and $s$, while primed objects refer to $\Lambda'$ and depend on
$v'$ and $s'$.

As shown by Georgi, Grinstein and Wise \cite{GGW}, the leading power
corrections to the infinite quark mass limit involve contributions of
two types.  The first come from terms in the expansion of the current
(\ref{Jexp1}) which involve operators containing a covariant
derivative. Their matrix elements can be parameterized as
\begin{equation}\label{zetamudef}
   \langle \Lambda' |\,\bar h'\,\Gamma^\alpha\,i D_\alpha\,h\,
   | \Lambda\rangle = \zeta_\alpha(v,v')\,\,\overline{\cal{U}}'\,
   \Gamma^\alpha\,{\cal{U}} \,.
\end{equation}
As in Ref.~I, we do not have to specify the nature of the matrix
$\Gamma^\alpha$ in the definition of the universal functions. At tree
level, however, $\Gamma^\alpha=\Gamma\,\gamma^\alpha$. Matrix elements
of operators containing a derivative acting on $h'$ are, as usual,
obtained from this by complex conjugation and interchange of the
velocity and spin labels. The most general decomposition of
$\zeta_\alpha$ involves two scalar functions defined by \cite{GGW}
\begin{equation}
   \zeta_\alpha(v,v') = \zeta_+(w)\,(v+v')_\alpha
   + \zeta_-(w)\,(v-v')_\alpha \,.
\end{equation}
As in the case of the mesons, one can use the equation of motion
$i v\!\cdot\!D h=0$ and the known spatial dependence (\ref{space}) of
the states in the effective theory to put constraints on these form
factors. One finds \cite{GGW}
\begin{eqnarray}
   \zeta_+(w) &=& {\bar\Lambda\over 2}\,{w-1\over w+1}\,\zeta(w)
    \,, \nonumber\\
   \zeta_-(w) &=& {\bar\Lambda\over 2}\,\zeta(w) \,.
\end{eqnarray}
 From these relations it follows that the matrix element in
(\ref{zetamudef}) vanishes at zero recoil.

The form factors also receive corrections from insertions of higher
order terms in the effective Lagrangian (\ref{lpower}) into matrix
elements of the lowest order current $J=\bar h'\,\Gamma\,h$. In fact,
the contribution of the chromo-magnetic operator vanishes by Lorentz
invariance, and the entire effect takes the form of a correction to the
Isgur-Wise function $\zeta(w)$:
\begin{equation}\label{Adef}
   \langle \Lambda' |\,i\!\int\!{\rm d}x\,T\big\{\,
   J(0),{\cal{L}}_1(x) \,\big\} \,| \Lambda \rangle
   = A(w)\,\,\overline{\cal{U}}'\,\Gamma\,\,{\cal{U}} \,.
\end{equation}

It is now straightforward to compute the form factors $F_i$ and $G_i$
at subleading order in HQET in terms of $\bar\Lambda$ and the universal
form factors $\zeta(w)$ and $A(w)$. Introducing the functions
\begin{eqnarray}
   {\cal{B}}_1(w) &=& \bar\Lambda\,{w-1\over w+1}\,\zeta(w) + A(w)
    \,, \nonumber\\
   {\cal{B}}_2(w) &=& - {2\bar\Lambda\over w+1}\,\zeta(w) \,,
\end{eqnarray}
the result becomes \cite{GGW}
\begin{eqnarray}
   F_1(w) &=& \zeta(w) + \bigg({1\over 2 m_Q} + {1\over 2 m_{Q'}}\bigg)
    \Big[ {\cal{B}}_1(w) - {\cal{B}}_2(w) \Big] \,, \nonumber \\
   G_1(w) &=& \zeta(w) + \bigg({1\over 2 m_Q} + {1\over 2 m_{Q'}}\bigg)
    \,{\cal{B}}_1(w) \,, \nonumber \\
   F_2(w) &=& \phantom{-} G_2(w) = {1\over 2 m_{Q'}}\,{\cal{B}}_2(w)
    \,,\nonumber\\
   F_3(w) &=& - G_3(w) = {1\over 2 m_Q}\,{\cal{B}}_2(w) \,.
\end{eqnarray}
For the subleading form factors, vector current conservation [cf.\
(\ref{CVC})] implies
\begin{equation}
   {\cal{B}}_1(1) = 0 \quad\Leftrightarrow\quad A(1) = 0 \,.
\end{equation}
Thus, at zero recoil all leading power corrections are determined in
terms of ${\cal{B}}_2(1)=-\bar\Lambda$, and in particular one finds
that $G_1(1)=1$ is not renormalized at this order \cite{GGW}.

\section{Second Order Power Corrections}
\label{sec:4}

We are now in a position to extend this analysis to include corrections
of order $1/m^2$ (from now on $m$ will designate a generic heavy quark
mass).  As in the case of the mesons, we must discuss separately three
classes of contributions:  corrections to the current, corrections to
the effective Lagrangian, and mixed corrections.  We shall take them
each in turn.

\subsection{\bf Second Order Corrections to the Current}

The effective operators appearing at second order in the
expansion of the current (\ref{Jexp1}) are all bilinear in the
covariant derivative, a property which remains true even if one goes
beyond tree level. It is thus sufficient to analyze the matrix element
\begin{equation}\label{psidef}
   \langle \Lambda'|\,\bar h'\,(-i\overleftarrow{D}_{\!\alpha})\,
   \Gamma^{\alpha\beta}\,i D_\beta\,h\,|\Lambda\rangle =
   \psi_{\alpha\beta}(v,v')\,\,\overline{\cal{U}}'\,
   \Gamma^{\alpha\beta}\,{\cal{U}} \,.
\end{equation}
Considering the complex conjugate of this equation leads immediately to
the relation $\psi_{\alpha\beta}(v,v')=\psi^*_{\beta\alpha}(v',v)$.
Decomposing the form factor into symmetric and antisymmetric parts,
$\psi_{\alpha\beta}=\case{1}/{2}[\psi^S_{\alpha\beta} +
\psi^A_{\alpha\beta}]$, we then write down the general decomposition
\begin{eqnarray}
   \psi_{\alpha\beta}^S(v,v') &=& \psi_1^S(w)\,g_{\alpha\beta}
    + \psi_2^S(w)\,(v+v')_\alpha (v+v')_\beta
    + \psi_3^S(w)\,(v-v')_\alpha (v-v')_\beta \,, \nonumber\\
   && \nonumber \\
   \psi_{\alpha\beta}^A(v,v')
    &=& \psi_1^A(w)\,(v_\alpha v'_\beta - v'_\alpha v_\beta) \,.
\end{eqnarray}
The equation of motion implies $v^\beta\psi_{\alpha\beta}=0$, yielding
\begin{eqnarray}
   \psi_1^S + (w+1)\,\psi_2^S - (w-1)\,\psi_3^S + w\,\psi_1^A &=& 0
    \,, \nonumber\\
   (w+1)\,\psi_2^S + (w-1)\,\psi_3^S - \psi_1^A &=& 0 \,.
\end{eqnarray}
As with the mesons, it is convenient to use an integration by parts to
relate (\ref{psidef}) to matrix elements of operators in which two
derivatives act on the same heavy quark field.  We find
\begin{equation}\label{partint}
   \langle \Lambda' |\,\bar h'\,\Gamma^{\alpha\beta}\,
   i D_\alpha i D_\beta\,h\,| \Lambda\rangle =
   \psi_{\alpha\beta}(v,v')\,\,\overline{\cal{U}}'\,
   \Gamma^{\alpha\beta}\,\,{\cal{U}} + \bar\Lambda\,(v-v')_\alpha\,
   \zeta_\beta(w)\,\,\overline{\cal{U}}'\,\Gamma^{\alpha\beta}\,\,
   {\cal{U}} \,.
\end{equation}
In particular, we define form factors for the matrix elements
\begin{eqnarray}\label{phis}
   \langle \Lambda'|\,\bar h'\,\Gamma\,(i D)^2 h\,| \Lambda\rangle
   &=& \phi_0(w)\,\,\overline{\cal{U}}'\,\Gamma\,\,{\cal{U}} \,,
    \nonumber \\
   \langle \Lambda' |\,\bar h'\,\Gamma^{\alpha\beta} G_{\alpha\beta}
   \,h\,|\Lambda\rangle &=& \phi_1(w)\,(v_\alpha v'_\beta -
    v'_\alpha v_\beta)\,\,\overline{\cal{U}}'\,\Gamma^{\alpha\beta}
    \,{\cal{U}} \,.
\end{eqnarray}
We may then use (\ref{partint}) and the relations given by the equation
of motion to write the form factors $\psi_i$ in terms of $\phi_i$,
$\zeta$, and $\bar\Lambda$:
\begin{eqnarray}\label{psirels}
   \psi^S_1 &=& \phi_0 + w\,\phi_1 + {w-1\over w+1}\,\bar\Lambda^2\,
    \zeta \,, \nonumber \\
   \psi^S_2 &=& -{1\over 2(w+1)} \bigg[ \phi_0 + (2w-1)\,\phi_1
    + {(2-w)(w-1)\over w+1}\,\bar\Lambda^2\,\zeta \bigg] \,,
    \nonumber \\
   \psi^S_3 &=& {1\over 2(w-1)} \Big[ \phi_0 + (2w+1)\,\phi_1 \Big]
    - {w\over 2(w+1)}\,\bar\Lambda^2\,\zeta \,, \nonumber \\
   \psi^A_1 &=& \phi_1 - {w-1\over w+1}\,\bar\Lambda^2\,\zeta \,,
\end{eqnarray}
where we omit the kinematic argument $w$ in the form factors. It
follows from (\ref{lamdef}) that the function $\phi_0(w)$ is normalized
at zero recoil, $\phi_0(1)=\lambda$. The equation of motion then
implies $\phi_1(1)=-\case1/3\lambda$. From the relations
(\ref{psirels}) we see that, as in the meson case, at zero recoil all
matrix elements of second order currents may be written in terms of the
single parameter $\lambda$, since
\begin{equation}
   \psi_{\alpha\beta}(v,v) = {\lambda\over 2}\,\big(g_{\alpha\beta}
   - v_\alpha v_\beta\big) \,.
\end{equation}
Furthermore, only the last operator in (\ref{Jexp1}) contributes at
zero recoil, yielding corrections of order $\lambda/m_Qm_{Q'}$.

\subsection{\bf Corrections to the Lagrangian}

We now turn to $1/m^2$ corrections which come from insertions of higher
dimension operators from the effective Lagrangian into matrix elements
of the lowest order current $J=\bar h'\,\Gamma\,h$. These fall into
three classes.  First, there are insertions of the second order
effective Lagrangian ${\cal L}_2$.  Although there are two new
operators at this order, only one of them gives a nonzero contribution.
This follows simply from Lorentz invariance, for the same reason that
the chromo-magnetic operator at order $1/m$ gave no contribution.  We
then define
\begin{equation}
   \langle \Lambda'|\,i\!\int\!{\rm d}x\,T\big\{\,J(0),{\cal{L}}_2(x)
   \,\big\} \,| \Lambda \rangle = Z_1\,B(w)\,\,
   \overline{\cal{U}}'\,\Gamma\,\,{\cal{U}} \,.
\end{equation}
Insertions of ${\cal L}'_2$ are parameterized by the same function.

Second, there are corrections which come from two insertions of the
first order correction ${\cal L}_1$.  These have the structure
\begin{eqnarray}
   \langle \Lambda' |\,\case{i^2}/{2}\!\int\!{\rm d}x{\rm d}y\,
   &T& \big\{\, J(0),{\cal{L}}_1(x),{\cal{L}}_1(y) \,\big\}
    \,| \Lambda \rangle \nonumber\\
   &=& C_1(w)\,\,\overline{\cal{U}}'\,\Gamma\,\,{\cal{U}}
    + Z^2\,C_{\alpha\beta\gamma\delta}(v,v')\,\,\overline{\cal{U}}'\,
    \Gamma\,P_+\,s^{\alpha\beta} P_+\,s^{\gamma\delta}\,{\cal{U}} \,,
\end{eqnarray}
where we decompose
\begin{eqnarray}\label{Cdecomp}
   C_{\alpha\beta\gamma\delta}(v,v')
   &=& C_2(w)\,(g_{\alpha\gamma}\,g_{\beta\delta} -
    g_{\alpha\delta}\,g_{\beta\gamma}) \nonumber \\
   &+& C_3(w)\,(g_{\alpha\gamma}v'_\beta v'_\delta
    -g_{\beta\gamma}v'_\alpha v'_\delta - g_{\alpha\delta}v'_\beta
    v'_\gamma + g_{\beta\delta}v'_\alpha v'_\gamma) \,.
\end{eqnarray}
The matrix elements for a double insertion of ${\cal L}'_1$ are given
by the same formula, but with $C_{\alpha\beta\gamma\delta}(v,v')$
replaced by $C_{\gamma\delta\alpha\beta}(v',v) =
C_{\alpha\beta\gamma\delta}(v',v)$.

Finally, there are corrections from an insertion of both ${\cal L}_1$
and ${\cal L}'_1$.  These have the structure
\begin{eqnarray}\label{Ddef}
   \langle \Lambda' |\,i^2\!\int\!{\rm d}x{\rm d}y\,
   &T& \big\{\, J(0),{\cal{L}}_1(x),{\cal{L}}'_1(y) \,\big\}
    \,| \Lambda \rangle \nonumber\\
   &=& D_1(w)\,\,\overline{\cal{U}}'\,\Gamma\,\,{\cal{U}}
    + ZZ' D_{\alpha\beta\gamma\delta}(v,v')\,\,\overline{\cal{U}}'\,
    s^{\alpha\beta} P'_+\,\Gamma\,P_+\,s^{\gamma\delta}\,{\cal{U}} \,.
\end{eqnarray}
We decompose $D_{\alpha\beta\gamma\delta}$ analogously to (\ref{Cdecomp}):
\begin{eqnarray}
   D_{\alpha\beta\gamma\delta}(v,v')
   &=& D_2(w)\,(g_{\alpha\gamma}\,g_{\beta\delta} -
    g_{\alpha\delta}\,g_{\beta\gamma}) \nonumber \\
   &&+ D_3(w)\,(g_{\alpha\gamma}v_\beta v'_\delta
    -g_{\beta\gamma}v_\alpha v'_\delta - g_{\alpha\delta}v_\beta
    v'_\gamma + g_{\beta\delta}v_\alpha v'_\gamma)
\end{eqnarray}
Note that $D_{\alpha\beta\gamma\delta}$ obeys the symmetry constraint
$D_{\alpha\beta\gamma\delta}(v,v')=D_{\gamma\delta\alpha\beta}(v',v)$.

\subsection{\bf Mixed Corrections to the Current and the Lagrangian}

Finally, we turn to second order corrections arising from insertions of
${\cal L}_1$ into matrix elements of first order corrections to the
current.  The structures of interest are
\begin{eqnarray}\label{Edef}
   \langle \Lambda' |\,i\!\int\!{\rm d}x\, &T& \big\{\,\bar
    h'\,\Gamma^\gamma\,i D_\gamma\,h,{\cal{L}}_1(x) \,\big\} \,
    |\Lambda\rangle \nonumber \\
   &=& E_\gamma(v,v')\,\,\overline{\cal{U}}'\,\Gamma^\gamma\,{\cal{U}}
    + Z E_{\gamma\alpha\beta}(v,v')\,\,\overline{\cal{U}}'\,
    \Gamma^\gamma P_+\,s^{\alpha\beta}\,{\cal{U}} \,, \nonumber \\
   && \\
    \langle \Lambda' |\,i\!\int\!{\rm d}x\, &T& \big\{\,
     \bar h'\,(-i\overleftarrow{D}_{\!\gamma})\,\Gamma^\gamma\,h,
     {\cal{L}}_1(x) \,\big\} \,|\Lambda\rangle \nonumber \\
   &=& E'_\gamma(v,v')\,\,\overline{\cal{U}}'\,\Gamma^\gamma\,{\cal{U}}
    + Z E'_{\gamma\alpha\beta}(v,v')\,\overline{\cal{U}}'\,
    \Gamma^\gamma P_+\,s^{\alpha\beta}\,{\cal{U}} \,. \nonumber
\end{eqnarray}
Again, insertions of ${\cal L}'_1$ give rise to the conjugate matrix
elements, with primed quantities interchanges with unprimed. We
parameterize
\begin{eqnarray}
   E_\gamma(v,v') &=& E_1(w)\,v_\gamma + E_2(w)\,v'_\gamma \,,
    \nonumber\\
   E'_\gamma(v,v') &=& E'_1(w)\,v_\gamma + E'_2(w)\,v'_\gamma \,,
    \nonumber\\
   && \\
   E_{\gamma\alpha\beta}(v,v') &=& E_3(w)\,(g_{\gamma\alpha} v'_\beta
    - g_{\gamma\beta} v'_\alpha) \,,\nonumber\\
   E'_{\gamma\alpha\beta}(v,v') &=& E'_3(w)\,(g_{\gamma\alpha} v'_\beta
    - g_{\gamma\beta} v'_\alpha) \,. \nonumber
\end{eqnarray}
The equation of motion implies $v^\gamma E_\gamma = v'^\gamma E'_\gamma
= 0$, yielding $E_1=-w\,E_2$ and $E'_2=-w\,E'_1$.  There are no
conditions on $E_3$ and $E'_3$.

As discussed in detail in Appendix~C of Ref.~I, the two matrix elements
in (\ref{Edef}) may be related to each other by an integration by
parts. Because there are fewer possible Lorentz structures for the
heavy baryons than for the mesons, here these relations take a
particularly simple form, namely
\begin{eqnarray}
   E_\gamma -  E'_\gamma &=& \bar\Lambda\,(v-v')_\gamma\,A
    + v_\gamma \big[ \phi_0 - \lambda\,\zeta \big] \,, \nonumber \\
   E_3 - E'_3 &=& 0 \,.
\end{eqnarray}
Hence we are left with only one new independent form factor, $E_3$. The
others may be written
\begin{eqnarray}
   E_1 &= - w E_2 &= {w\over w+1} \Big[ w\,\widetilde{\phi}
    + \bar\Lambda\,A \Big] \,, \nonumber \\
   E'_2 &= - w E'_1 &= {w\over w+1} \Big[ -\widetilde{\phi}
    + \bar\Lambda\,A \Big] \,,
\end{eqnarray}
where
\begin{equation}\label{tilphi}
   \widetilde\phi(w) = {\phi_0(w) - \lambda\,\zeta(w)\over w-1}
\end{equation}
is a nonsingular function as $w\to 1$, since $\phi_0(1)=\lambda$.

Finally, we note that the equations of motion imply that the form
factor $E_\gamma$ takes the form $E_\gamma = E_1\,
(v_\gamma-w\,v'_\gamma)$, which vanishes as $v\to v'$. The expression
for $E'_\gamma$ has a similar structure, while the kinematic structures
multiplying $E_3$ and $E'_3$ vanish at zero recoil. Hence, as with the
mesons, the mixed corrections give no contribution at zero recoil to
form factors which are not kinematically suppressed.

\subsection{\bf Form Factors and Normalization Conditions}

We have introduced a set of ten new universal functions which describe
the $1/m^2$ corrections to heavy $\Lambda$ baryon form factors in the
heavy quark expansion. Two of these, $\phi_0$ and $\phi_1$,
parameterize the corrections to the current, seven more, $B$, $C_i$
and $D_i$, for $i=1,2,3$, parameterize the effects of higher order
terms in the effective Lagrangian, and one, $E_3$, is needed in
order to include mixed corrections to the current and the Lagrangian.
It is now straightforward to express the vector and axial vector form
factors $F_i$ and $G_i$ up to order $1/m^2$ in terms of these universal
functions. To this end it is useful, as in the meson case, to collect
certain combinations of universal form factors by introducing the
functions
\begin{eqnarray}
   b_1 &=& \lambda\,\zeta + B + C_1 - 3 C_2 + 2(w^2-1) C_3 \nonumber\\
    &&+ (w-1)\,(\phi_1-2 E_3) + {w-1\over w+1}\,
    (w\widetilde{\phi} + \bar\Lambda A) \,, \nonumber\\
   b_2 &=& - 2\,(\phi_1-2 E_3) - {2\over w+1}\,
    (w\widetilde{\phi} + \bar\Lambda A) \,, \nonumber\\
   b_3 &=& D_1 + D_2 - \phi_1 + {w-1\over w+1}\,
    \Big[ \bar\Lambda^2\zeta - 2\,(\widetilde{\phi} - \bar\Lambda A)
    \Big] \,, \nonumber\\
   b_4 &=& {4\over w+1}\,(\widetilde{\phi} - \bar\Lambda A)
    \,, \nonumber\\
   b_5 &=& -2 D_2 - 2(w-1) D_3
    - 3 {w-1\over(w+1)^2}\,\bar\Lambda^2\zeta \nonumber\\
    && + {1\over w+1}\,\Big[ - \phi_0 + (2-w)\,\phi_1
    + 2\,(\widetilde{\phi} - \bar\Lambda A) \Big] \,, \nonumber\\
   b_6 &=& 2 D_2 + 2(w+1) D_3 - {\bar\Lambda^2\over w+1}\,\zeta
    \nonumber\\
    && + {1\over w-1}\,\big[ \phi_0 + (2+w)\,\phi_1 \big]
    + {2\over w+1}\,(\widetilde{\phi} - \bar\Lambda A) \,.
\end{eqnarray}
Note that the term $\lambda\zeta$ in $b_1$ arises from substituting the
relation (\ref{spinors}) between the physical baryon spinors $u(v,s)$,
which appear in the definition of the form factors $F_i$ and $G_i$, and
the effective spinors ${\cal{U}}(v,s)$ of HQET, into the leading order
matrix elements (\ref{lowest}). Let us furthermore specialize to
transitions of the type $\Lambda_b\to\Lambda_c$, and abbreviate
$\varepsilon_b= 1/2 m_b$ and $\varepsilon_c=1/2 m_c$. We then find
\begin{eqnarray}\label{finalres}
   F_1 &=& \zeta + (\varepsilon_c + \varepsilon_b)\,
    \big[ {\cal{B}}_1 - {\cal{B}}_2 \big]              
    + (\varepsilon_c^2 + \varepsilon_b^2)\,
    \big[ b_1 - b_2 \big] + \varepsilon_c \varepsilon_b\,
    \big[ b_3 - b_4 \big] \,, \nonumber\\
   F_2 &=& \varepsilon_c\,{\cal{B}}_2 + \varepsilon_c^2\,b_2
    + \varepsilon_c \varepsilon_b\,b_5 \,, \nonumber\\
   F_3 &=& \varepsilon_b\,{\cal{B}}_2 + \varepsilon_b^2\,b_2
    + \varepsilon_c \varepsilon_b\,b_5 \,, \nonumber\\
   && \\
   G_1 &=& \zeta + (\varepsilon_c + \varepsilon_b)\,{\cal{B}}_1
    + (\varepsilon_c^2 + \varepsilon_b^2)\,b_1
    + \varepsilon_c \varepsilon_b\,b_3 \,, \nonumber\\
   G_2 &=& \phantom{-} \varepsilon_c\,{\cal{B}}_2 + \varepsilon_c^2\,
    b_2 + \varepsilon_c \varepsilon_b\,b_6 \,, \nonumber\\
   G_3 &=& - \varepsilon_b\,{\cal{B}}_2 - \varepsilon_b^2\,b_2
    - \varepsilon_c \varepsilon_b\,b_6 \,. \nonumber
\end{eqnarray}
Recall that $\widetilde\phi$ was defined in terms of other universal
functions in (\ref{tilphi}).

Order by order in the heavy quark expansion, the normalization
condition (\ref{CVC}) imposes a constraint on the universal functions
of HQET. Hence, in addition to $\zeta(1)=1$ and $A(1)=0$, there is
a relation at zero recoil between the form factors which arise at order
$1/m^2$. Evaluating the sum of $F_i$ for equal masses, we obtain
\begin{equation}
   2 b_1(1) + b_3(1) - b_4(1) + 2 b_5(1) = 0 \,,
\end{equation}
which is equivalent to
\begin{equation}\label{normal}
   2 B(1) + 2 C_1(1) + D_1(1) - 6 C_2(1) - 3 D_2(1) = - \lambda \,.
\end{equation}

\section{Applications to Semileptonic $\Lambda_{\lowercase {b}}$ Decays}
\label{sec:5}

In this section we apply our results to semileptonic $\Lambda_b$ decays
and give some estimates of the size of the second order corrections.
For simplicity, and in order to focus on what is new in our analysis,
we shall continue to ignore radiative corrections.

\subsection{\bf $\Lambda_b\to\Lambda_c\,\ell\,\nu$ Decays Near Zero
Recoil}

The semileptonic decay $\Lambda_b\to\Lambda_c\,\ell\,\nu$ is
particularly simple to analyze near the zero recoil point $w=1$, where
the invariant mass $q^2$ of the lepton pair takes on its maximum value
$q^2_{\rm max}=(m_{\Lambda_b}-m_{\Lambda_c})^2$.  In the limit of
vanishing lepton mass, angular momentum conservation requires that the
weak matrix element $\langle\Lambda_c(v,s')\,|\,(V^\mu-A^\mu)\,|\,
\Lambda_b(v,s)\rangle$ depend only on the function $G_1(1)$. The
differential decay rate near this point is given by
\begin{equation}
   \lim_{w\to1}\,{1\over\sqrt{w^2-1}}\,
   {{\rm d}\Gamma(\Lambda_b\to\Lambda_c\,\ell\,\nu)\over{\rm d}w}\,
   ={G_F^2\,|\,V_{cb}|^2\over 4\pi^3}\,m_{\Lambda_c}^3\,
   (m_{\Lambda_b}-m_{\Lambda_c})^2\,|G_1(1)|^2 \,.
\end{equation}
The form factor $G_1(1)$ is protected against corrections at order
$1/m$ \cite{GGW}, but it receives contributions from order $1/m^2$
effects.  Incorporating the normalization condition (\ref{normal}), we
find
\begin{equation}\label{G1at1}
   G_1(1) = 1 + (\varepsilon_c-\varepsilon_b)^2\,b_1(1)
   +\varepsilon_c\varepsilon_b \big[ b_4(1) - 2 b_5(1) \big] \,.
\end{equation}
We may estimate the size of the corrections to $G_1(1)$ by considering
the form of the corresponding vector current matrix element at zero
recoil, given by
\begin{equation}
   \langle\Lambda_c(v,s')\,|\,V^\mu\,|\,\Lambda_b(v,s)\rangle
   = 2 \sqrt{m_{\Lambda_c} m_{\Lambda_b}}\,F(1)\,v^\mu \,,
\end{equation}
where
\begin{equation}
   F(1) \equiv \sum_{i=1,2,3}\,F_i(1) = 1 +
   (\varepsilon_c-\varepsilon_b)^2\,b_1(1) \,.
\end{equation}
The function $F(1)$ measures the overlap of the wavefunctions of the
light degrees of freedom between a $\Lambda_b$ and a $\Lambda_c$
baryon.  While
the light quarks and gluons were insensitive to the mass of the heavy
quark in the strict $m\to\infty$ limit and in precisely the same
configuration in a $\Lambda_b$ and a $\Lambda_c$, at order $1/m^2$ the
wavefunctions differ from each other and the overlap is incomplete
($F(1)<1$).  We may estimate the size of this difference in a
nonrelativistic model in which a $\Lambda_Q$ baryon is composed of a
constituent diquark of mass $m_{qq}\approx\bar\Lambda\approx 700\,
{\rm MeV}$, orbiting about the heavy quark.  In this case the
$m_Q$-dependence of the overlap integral comes from the
$m_Q$-dependence of the reduced mass $m_{qq}^{\rm red} =
m_{qq}m_Q/(m_{qq}+m_Q)$ of the diquark.  We then obtain the
estimate
\begin{equation}
   b_1(1)\approx -3\bar\Lambda^2\approx -1.5\,{\rm GeV^2} \,.
\end{equation}
This combination is the same as appears in the first term which
corrects $G_1(1)$.

The second term, $b_4(1)-2b_5(1)=\case{4}/{3}\lambda+4D_2(1)$, is
harder to estimate.  However, we note that the function $D_2$ arises
from the double insertion of the chromo-magnetic operator in
${\cal L}_1$, and there are indications from QCD sum rules that it is
likely to be quite small \cite{Sublea}. Furthermore, for heavy mesons
sum rules predict a value for the analog of $\lambda$ which is positive
and about 1 GeV \cite{SR2}. Let us for the sake of argument assume such
a value here. Using $m_c=1.5$ GeV and $m_b=4.8$ GeV, we then we obtain
\begin{equation}
   G_1(1)\approx 1 - 7.7\% + 4.6\% \,.
\end{equation}
While this estimate is of course quite rough, it is reasonable to
expect at least that the signs of the two terms are as we claim, such
that there is a partial cancellation of the two contributions. Then if
the magnitudes are even approximately correct, one may argue that
$1/m^2$ corrections to $G_1(1)$ at the level of ten percent would be
surprising. Consequently, we expect the semileptonic decay
$\Lambda_b\to\Lambda_c\,\ell\,\nu$ to be well described by HQET.

\subsection{\bf Asymmetry Parameters in $\Lambda_b\to
\Lambda_c\,\ell\,\nu$ Decays}

The angular distributions in the cascade $\Lambda_b\to\Lambda_c\,
\ell\,\nu \nu \to\Lambda\,X\,\ell\,\nu$ provide an efficient analysis
of polarization effects in semileptonic $\Lambda_b$ decay. This is
particularly true at the kinematic endpoint $q^2=0$, where only the
helicity amplitudes in which a longitudinal virtual $W$ boson is
emitted contribute.  Such effects are discussed at length in
Ref.~\cite{KoKr}, to which we refer the interested reader for details.
Here we shall merely cite the final expressions.

There are several asymmetry parameters which are particularly
interesting at $q^2=0$ within the heavy quark expansion.  The simplest
comes from the distribution in the angle $\theta_\Lambda$ between the
$\Lambda$ and $\Lambda_c$ directions.  The differential decay width in
this variable is given by
\begin{equation}
   {{\rm d}\Gamma\over{\rm d}q^2\,{\rm d}\cos\theta_\Lambda}\propto
   1 + \alpha\,\alpha_{\Lambda_c}\cos\theta_\Lambda \,,
\end{equation}
where $\alpha$ is the asymmetry parameter of the $\Lambda_b$ decay, and
$\alpha_{\Lambda_c}$ is the measured asymmetry parameter in the decay
$\Lambda_c\to\Lambda\,X$. For the nonleptonic decay $\Lambda_c^+\to
\Lambda\,\pi^+$, a particularly useful mode, there are recent
measurements $\alpha_{\Lambda_c}=-1.0^{+0.4}_{-0.0}$ \cite{CLEO} and
$\alpha_{\Lambda_c}=-0.96\pm0.42$ \cite{ARGUS}.

Two additional asymmetry parameters which have interesting HQET
expansions may be defined for the decay of polarized $\Lambda_b$
baryons.  Let $P$ be the degree of polarization of the $\Lambda_b$, and
$\theta_P$ the angle between the $\Lambda_b$ polarization and the
direction of the $\Lambda_c$.  Then the parameter $\alpha_P$ is defined
by the form of the differential distribution,
\begin{equation}
   {{\rm d}\Gamma\over{\rm d}q^2\,{\rm d}\cos\theta_P}\propto
   1 - \alpha_P P\cos\theta_P \,.
\end{equation}
Further, let $\chi_P$ be the angle between the plane of the $\Lambda_c$
decay and the plane formed by the $\Lambda_b$ polarization and the
$\Lambda_c$ direction.  Then the angular distribution in $\chi_P$ is
given by
\begin{equation}
   {{\rm d}\Gamma\over{\rm d}q^2\,{\rm d}\chi_P}\propto
   1 - \gamma_P\,{\pi^2\over16}\,P\alpha_{\Lambda_c}\cos\chi_P \,,
\end{equation}
where $\gamma_P$ yet is another asymmetry parameter.

At $q^2=0$, the expressions for $\alpha$, $\alpha_P$ and $\gamma_P$ take
simple forms,
\begin{eqnarray}
   \alpha &=& - \alpha_P =
   - {1-|\,\epsilon\,|^2\over 1+|\,\epsilon\,|^2} \,, \nonumber\\
   \gamma_P &=& {2\,{\rm Re}(\epsilon)\over 1 + |\,\epsilon\,|^2} \,,
\end{eqnarray}
where
\begin{equation}
   \epsilon = {f_1(0) - g_1(0)\over f_1(0) + g_1(0)} \,.
\end{equation}
At leading order in HQET, this ratio vanishes since $f_1=g_1=\zeta$. In
this limit the asymmetries are predicted to be $\alpha=-\alpha_P=-1$
and $\gamma_P=0$ \cite{KoKr}. Using (\ref{fsandgs}) and
(\ref{finalres}), we find that there are no $1/m$ corrections to these
predictions. The leading power correction comes at order $1/m^2$:
\begin{equation}
   \epsilon_{1/m^2} = {1\over 4\zeta(w)} \Big\{
   (\varepsilon_c-\varepsilon_b)^2 \Big[ b_5(w) - b_6(w) \Big]
   + 2\varepsilon_c\varepsilon_b \Big[ 2 b_5(w) - b_4(w) \Big]
   \Big\} \,,
\end{equation}
where $w=(m_{\Lambda_b}^2+m_{\Lambda_c}^2)/2 m_{\Lambda_b}
m_{\Lambda_c}$ corresponding to $q^2=0$. Based on our previous
estimates we expect $\epsilon_{1/m^2}$ to be of the order of a few
percent. A contribution of similar magnitude comes from perturbative
corrections to the heavy quark currents at leading order in HQET. It is
given by \cite{Gave,QCD}
\begin{equation}
   \epsilon_{\rm QCD} = - {2\alpha_s\over 3\pi}\,
   {m_b m_c\over m_b^2 - m_c^2}\,\ln{m_b\over m_c} \approx -2.4\% \,,
\end{equation}
where we have used $\alpha_s/\pi=0.09$.

In view of its smallness, it will be virtually impossible to determine
$|\,\epsilon\,|$ from a measurement of $\alpha$ or $\alpha_P$, since
these parameters depend only on $|\,\epsilon\,|^2$ and should,
therefore, be very close to the asymptotic values given above. A
measurement of a nonzero asymmetry $\gamma_P$, on the other hand, would
provide a direct determination of Re$(\epsilon)$ and could yield
valuable information about the size of $1/m^2$ corrections.

\section{Matrix Elements of Excited Baryons}
\label{sec:6}

The entire analysis presented here could be extended to matrix elements
involving excited baryons, in particular to baryons of higher spin.  To
order $1/m$, this was done by Mannel, Roberts and Ryzak \cite{Robe}.
The $\Lambda_Q$ baryons which we have been considering are extremely
simple, because the light degrees of freedom are in a state of zero
total angular momentum, and hence the polarization of the baryon is the
same as the polarization of the heavy quark.  There is, however, an
excited state in which the spins of the light quarks are aligned so
that the light degrees of freedom have angular momentum $s_\ell=1$.
When combined with the heavy quark, this state becomes a degenerate
doublet of an excited spin-$\case1/2$ baryon, the $\Sigma_Q$, and a
spin-$\case3/2$ baryon, the $\Sigma^*_Q$.  The analysis of the
semileptonic decays of and into these states is analogous to that for
the mesons and $\Lambda_Q$ baryons, except that the states have to be
represented differently, and the counting of form factors is modified
accordingly. Rather than elaborate the entire analysis yet again, we
shall simply indicate how it differs from the cases already presented.

As for the pseudoscalar and vector mesons, it is convenient to assemble
the degenerate doublet $(\Sigma_Q,\,\Sigma_Q^*)$ into a single object.
This allows us to implement the spin symmetries in a compact formalism.
Let us represent the spin-$\case1/2$ $\Sigma_Q$ by the spinor
$\psi$ and the spin-$\case3/2$ $\Sigma_Q^*$ by the Rarita-Schwinger
vector-spinor $\psi^\mu$. In the heavy quark limit, these objects
satisfy $\rlap/v\,\psi=\psi$, $\rlap/v\,\psi^\mu= \psi^\mu$,
$v_\mu\psi^\mu=\gamma_\mu\psi^\mu=0$. Then the doublet is represented
by \cite{MRR,AFF}
\begin{equation}
   \Psi^\mu=\psi^\mu+\case1/{\sqrt3}(\gamma^\mu+v^\mu)\,
   \gamma^5\,\psi\,,
\end{equation}
which satisfies the constraints $v_\mu\Psi^\mu=0$ and $\rlap/v\,
\Psi^\mu=\Psi^\mu$.  It is straightforward to construct the analogs of
$\Psi^\mu$ for baryons of arbitrary spin \cite{AFF}.

 From here on the heavy quark expansion proceeds almost exactly as
before. For example, for semileptonic transitions of the form
$\Lambda_Q\to\Sigma_{Q'}$ or $\Lambda_Q\to\Sigma_{Q'}^*$, one repeats
the analysis of Secs.~\ref{sec:3} and \ref{sec:4}, but with an
additional index $\mu$ on all universal form factors.  There is,
however, a subtlety which must be considered.  The spin-parity
$s_\ell^P$ of the light degrees of freedom may be either in the series
$0^+,1^-,2^+,\dots$, in which case it is ``natural'', or in the series
$0^-,1^+,2^-,\dots$, in which case it is ``unnatural''.  As noted in
Ref.~\cite{Poli}, there are additional restrictions on the universal
functions which describe the transitions between ``natural'' and
``unnatural'' baryons \cite{MRR,AFF}. These restrictions may be imposed
\cite{Robe} by constructing form factors which are pseudotensors,
rather than tensors.

In particular, the $\Lambda_Q$ is a ``natural'' baryon, while the
$\Sigma_Q$ and $\Sigma_Q^*$ are ``unnatural''.  Hence at leading order,
$\Sigma\to\Sigma$ transitions are governed by a tensor form factor of
the form
\begin{eqnarray}
   \langle\Sigma'|\,\bar h'\,\Gamma\,h\,|\Sigma\rangle
   &=& K_{\mu\nu}(v,v')\,{\overline\Psi'}\vphantom{\Psi}^\mu\,\Gamma\,
    \Psi^\nu \nonumber\\
   &=& \Big[ K_1(w)\,g_{\mu\nu}+K_2(w)\,v_\mu v'_\nu \Big]\,
    {\overline\Psi'}\vphantom{\Psi}^\mu\,\Gamma\,\Psi^\nu \,,
\end{eqnarray}
while the leading $\Lambda\to\Sigma$ transitions would require a
pseudovector form factor. However,
\begin{equation}
   \langle\Sigma'|\,\bar h'\,\Gamma\,h\,|\Lambda\rangle =
   K_\mu(v,v')\,{\overline\Psi'}\vphantom{\Psi}^\mu\,\Gamma\,u = 0 \,,
\end{equation}
since no such object $K_\mu$ can be built from the available vectors
$v$ and $v'$.

Once this subtlety has been taken into account, the construction of the
heavy quark expansion proceeds just as before. Order by order, one
identifies the (pseudo) tensor-valued functions which describe a given
type of correction, performs a general decomposition in terms of
velocities to obtain the complete list of universal functions, and then
writes the physical matrix elements in terms of them.  The restrictions
imposed by the heavy quark spin symmetries are built into the formalism
from the start.  For example, one might consider the corrections to
$\Sigma\to\Sigma'$ transitions which arise from insertions of the first
order corrections to the effective Lagrangian. One finds five form
factors $L_i$, defined by
\begin{eqnarray}
   \langle \Sigma' |\,i\!\int\!{\rm d}x\, &T& \big\{\,
    J(0), \big[ h\,(iD)^2 h \big]_x \,\big\}\,| \Sigma \rangle
    \nonumber\\
   &=& \Big[ L_1(w)\,g_{\mu\nu} + L_2(w)\,v_\mu v'_\nu \Big] \,
    {\overline\Psi'}\vphantom{\Psi}^\mu\,\Gamma\,\Psi^\nu \,,
    \nonumber \\
   && \\
   \langle \Sigma' |\,i\!\int\!{\rm d}x\, &T& \big\{\,
    J(0), \big[ \bar h\,s^{\alpha\beta} G_{\alpha\beta} h \big]_x
    \,\big\}\,|\Sigma\rangle \nonumber\\
   &=& \Big[ L_3(w)\,(g_{\alpha\mu} g_{\beta\nu} - g_{\alpha\nu}
    g_{\beta\mu}) + L_4(w)\,(g_{\alpha\nu} v'_\beta v_\mu -
    g_{\beta\nu} v'_\alpha v_\mu) \nonumber \\
   &&+ L_5(w)\,(g_{\alpha\mu} v'_\beta v'_\nu
    - g_{\beta\mu} v'_\alpha v'_\nu) \Big] \,
    {\overline\Psi'}\vphantom{\Psi}^\mu\,\Gamma\,P_+\,s^{\alpha\beta}
    \Psi^\nu \,. \nonumber
\end{eqnarray}
This procedure clearly becomes more tedious as the spin of the baryons
increases and with higher order in the $1/m$ expansion; however, the
enumeration of form factors is straightforward, systematic and complete.

Finally, we note that in the case of $b\to c$ weak decays, it is only
the transitions of the form $\Lambda_b\to\Lambda_c,\,\Sigma_c,\,
\Sigma^*_c,\ldots$ which are likely to be of experimental interest.
This is because the excited bottom baryons will decay strongly (if the
mass splitting is sufficient to allow pion emission) or
electromagnetically to the ground state $\Lambda_b$, and thus their
weak decays will not be observable. On the other hand, the decays
$\Lambda_b\to\Sigma_c,\Sigma_c^*$ will be particularly interesting,
since they arise solely due to effects of order $1/m_c$ and higher.

\section{Summary}
\label{sec:7}

We have extended the analysis of $1/m^2$ corrections in the heavy quark
effective theory to the heavy baryons.  We have focused in detail on
the simplest case, the weak matrix elements relevant to the decay of a
heavy $\Lambda_Q$ to a heavy $\Lambda_{Q'}$.  Due to the trivial
Lorentz structure of the light degrees of freedom in this system, the
description of the power corrections is considerably simpler than for
the heavy mesons.  At order $1/m^2$, one needs a set of ten new
$m_Q$-independent Isgur-Wise functions of the kinematic variable
$v\cdot v'$, and a single new dimensionful parameter $\lambda$. Vector
current conservation forces a certain combination of form factors to
vanish at zero recoil.

We have given a rough estimate of the size of the second order
corrections for the semileptonic decay $\Lambda_b\to\Lambda_c\,
\ell\,\nu$.  We find a partial cancellation of $1/m^2$ corrections at
zero recoil, with the conclusion that large deviations from the
infinite quark mass limit are unlikely, and the heavy quark expansion
is well under control. Investigating briefly the asymmetry parameters
which may be defined in this decay, we have suggested a particular
measurement which would probe the $1/m^2$ corrections directly.
Finally, we have sketched the extension of the formalism to excited
heavy baryons of arbitrary spin.

\acknowledgements
M.N. gratefully acknowledges financial support from the BASF
Aktiengesellschaft and from the German National Scholarship Foundation.
This work was supported by the Department of Energy, contract
DE-AC03-76SF00515.

\end{document}